\documentclass[floatfix,aps,amsmath,nofootinbib,onecolumn,10pt]{revtex4}

\usepackage{amsmath}
\usepackage{graphicx}
\usepackage{bm}
\usepackage{rotating}
\usepackage{array}
\usepackage{float}

\def\({\left(}
\def\){\right)}
\def\[{\left[}
\def\]{\right]}

\def\e{\begin{equation}}
\def\q{\end{equation}}
\def\m{\begin{eqnarray}}
\def\n{\end{eqnarray}}

\makeatletter
\renewcommand\@bibitem[1]{\item\if@filesw \immediate\write\@auxout
       {\string\bibcite{#1}{\the\value{\@listctr}}}\fi\ignorespaces}
\makeatother

\begin{document}

\title{Constraints on Phenomenological Amplitudes of CMB Anisotropy with Multi-Datasets}

\author{Yang Han, Lu Chen\footnote{Corresponding Author: chenlu@sdnu.edu.cn}, Guo-Hao Li, Pei-Yuan Xu }
\affiliation{ Shandong Provincial Key Laboratory of Light Field Manipulation Physics and Applications \\ $\&$ School of Physics and Optoelectronics, \\Shandong Normal University, Jinan 250358, China\\} 
\date{\today}

\begin{abstract}
\noindent 
Cosmic microwave background anisotropies encode crucial information about the early Universe and fundamental cosmological physics. Although the standard $\Lambda$CDM model provides a successful description of cosmic evolution, persistent cosmological tensions and subtle small-scale anomalies still challenge its internal consistency.
In this paper, we investigate six phenomenological amplitude parameters $A_{\rm new}$ (new$=$L, SW, Dop, eISW, lISW, Pol) corresponding to the key effects related to CMB anisotropy---the Lensing, Sachs–Wolfe, Doppler, early Integrated Sachs–Wolfe, late Integrated Sachs–Wolfe, and Polarization effects, respectively. Using modified CAMB and Cobaya packages, we constrain the $\Lambda$CDM$+A_{\rm new}$ models with two data combinations---Planck+DESI+PantheonPlus (PDP) and Planck+ACT+DESI+PantheonPlus (PADP). 
Only the $\Lambda$CDM+$A_{\rm{L}}$ is favored by AIC, with $A_{\rm{L}}=1.0656_{-0.0303}^{+0.0304}$ from PDP and $A_{\rm{L}}=1.0795_{-0.0289}^{+0.0260}$ from PADP, which implies 2.16$\sigma$ and 3.06$\sigma$ deviation from the $\Lambda$CDM model;
values of $A_{\rm{SW}}$ show 1.21$\sigma$ and 1.96$\sigma$ deviations to 1;
$A_{\rm{lISW}}$ is poorly constrained because the lISW effect has negligible influence at $\ell \geq$30;
and others are consistent with the $\Lambda$CDM model.
Moreover, no noticeable improvement on the Hubble and $\sigma_8$ tensions is found within these one-parameter extended scenarios.
ACT DR6 high-$\ell$ data strengthens the $\Lambda$CDM$+A_{\rm{L}}$ preference over the $\Lambda$CDM model, and reduces $A_{\rm Pol}$ uncertainty by more than one order of magnitude, highlighting the importance of ground-based high-$\ell$ observations for future CMB analyses.

\end{abstract}

\pacs{???}

\maketitle


\section{Introduction} 
\label{sec:int}

The Cosmic Microwave Background (CMB) serves as one of the most robust observational tools for investigating the fundamental properties of our universe.
Since its discovery by Penzias and Wilson in 1965~\cite{Penzias:1965wn}, it has undergone increasingly precise measurements, from the COBE satellite's first detection of temperature anisotropies~\cite{COBE:1992syq}, through WMAP's full-sky mapping~\cite{WMAP:2003ivt} to the Planck satellite's high-precision measurement of temperature and polarization power spectra~\cite{Planck:2018vyg,Planck:2019nip}.
In particular, the systematic monitoring of CMB anisotropies, encompassing temperature and polarization variations, has become a cornerstone of modern cosmological research, enabling the validation and refinement of cosmological models.
During decades of research, people have established a standard cosmological model---the $\Lambda$CDM model, which describes a universe mainly dominated by cold dark matter and cosmological constant $\Lambda$~\cite{Efstathiou:1990xe}, with a six-parameter framework.
It has been validated by many measurements, yet it remains critical to ensure its internal consistency by examining the individual physical processes that contribute to CMB anisotropies, especially given that the standard model still faces unresolved issues, such as the Hubble and $\sigma_8$ tensions.

The CMB angular power spectrum, a key observable in cosmological measurement, is shaped by multiple physical mechanisms, each contributing uniquely to its structure.
The temperature anisotropies of the CMB originate from various physical mechanisms in the process of cosmic evolution. At large angular scales, the Sachs-Wolfe (SW) effect dominates, representing the gravitational redshift or blueshift experienced by CMB photons as they depart from potential wells at the last scattering surface~\cite{Sachs:1967er}. 
At the medium scales, the Doppler effect from baryon velocity and the early Integrated Sachs-Wolfe (eISW) effect from time-evolving potentials during radiation-matter equality contribute significantly~\cite{Hu:1997hp,Zaldarriaga:1995gi}. 
The late Integrated Sachs-Wolfe (lISW) effect, generated by the decay of gravitational potentials due to dark energy, provides a unique signature of cosmic acceleration at low-$\ell$~\cite{Rees:1968zza}. 
Additionally, the Polarization effect, generated by anisotropic Thomson scattering of photons by electrons,  provides information about the primordial perturbations and the reionization history~\cite{Hu:1997hp,Zaldarriaga:1996xe}. 
Gravitational lensing by large scale structure further modifies the CMB power spectra by smoothing the acoustic peaks at small scales and converting E-mode polarization into B-mode polarization~\cite{Lewis:2006fu,Planck:2013mth}.

Despite the remarkable success of the $\Lambda$CDM model, several intriguing tensions and anomalies have emerged in recent CMB observations. The Hubble tension, a persistent discrepancy between the locally measured Hubble constant $H_0 \approx 73\, \text{km s}^{-1}\text{Mpc}^{-1}$~\cite{Riess:2021jrx} and the CMB inferred value $H_0 \approx 67.4\, \text{km s}^{-1}\text{Mpc}^{-1}$~\cite{Planck:2018vyg} has motivated extensive searches for new physical mechanisms beyond the standard model. 
The $\sigma_8$ tension, which describes a significant discrepancy about 2$\sigma$ between the CMB predicted amplitude of matter fluctuations and lower values inferred from late-time large-scale structure ~\cite{Abdalla:2022yfr}, has become another major challenge to the $\Lambda$CDM model.
Furthermore, the Planck data exhibit a mild preference for an enhanced Lensing effect, parameterized by $A_{\rm{L}} > 1$~\cite{Planck:2018vyg,DiValentino:2020zio}.
This lensing anomaly has been investigated in many works, with some suggesting it may indicate a genuine physical effect while others attribute it to statistical fluctuations or systematic effects~\cite{DiValentino:2021izs,Addison:2023fqc}.

To systematically investigate the contributions of different physical effects to the CMB power spectra and test the consistency of the $\Lambda$CDM model, phenomenological amplitude parameters have been introduced as a model-independent approach. 
By rescaling the contributions of individual physical effects, we can quantify the agreement between observations and theoretical predictions for each physical effect separately~\cite{Planck:2015mrs,Planck:2018lbu,Planck:2018vyg,Planck:2019nip}. 
Previous studies have employed similar phenomenological amplitude parameters to test the consistency of the $\Lambda$CDM model, but these analyses have largely focused on Planck data, without incorporating high-resolution ground-based observations.
Ground-based CMB experiments, such as the Atacama Cosmology Telescope (ACT)~\cite{Naess:2020wgi,AtacamaCosmologyTelescope:2013swu}, offer unique advantages in constraining small-scale CMB features, which are critical for resolving lensing effects and polarization signals. The ACT Data Release 6 (DR6)~\cite{ACT:2023kun,ACT:2023dou,AtacamaCosmologyTelescope:2025blo} provides high-sensitivity measurements of CMB at small angular scales, complementing the large-scale coverage of Planck and enabling more robust constraints on phenomenological parameters.
Combined with Baryon Acoustic Oscillation (BAO) measurements from the Dark Energy Spectroscopic Instrument (DESI) Data Release 2 (DR2)~\cite{DESI:2025zgx,DESI:2025zpo} and SNIa data from the PantheonPlus sample~\cite{Scolnic:2021amr}, these datasets may enable better constraints on cosmological parameters and phenomenological amplitudes.
In this work, we perform a systematic analysis of six phenomenological amplitude parameters, corresponding to the SW effect ($A_{\rm{SW}}$), Doppler effect ($A_{\rm{Dop}}$), eISW effect ($A_{\rm{eISW}}$), lISW effect ($A_{\rm{lISW}}$), Polarization effect ($A_{\rm{Pol}}$), and Lensing effect ($A_{\rm{L}}$), within the $\Lambda$CDM framework. 
Then we constrain these parameters using two dataset combinations: Planck$+$DESI$+$PantheonPlus (named PDP hereafter) and Planck$+$ACT$+$DESI$+$PantheonPlus (denoted as PADP). 

This paper is organized as follows. In Sec.~\ref{sec:the}, we present the theoretical framework, introducing the definitions of the phenomenological amplitude parameters. In Sec.~\ref{sec:res}, we describe the datasets, methodology, and our results. Finally, in Sec.~\ref{sec:sum}, we summarize our findings and discuss their implications for understanding the physical mechanisms of CMB anisotropies and the consistency of the $\Lambda$CDM model.

\section{Theory}
\label{sec:the}
In this section, we modify the equations for calculating the CMB power spectrum to demonstrate the contributions of different physical effects following previous works.  

During their journey from the last scattering surface to our telescopes, CMB photons are perturbed by the gravitational fields, causing their trajectories to deflect, which is known as the Lensing effect. Such deflection alters the spatial distribution of the observed temperature and polarization patterns, leaving identifiable imprints on the angular power spectrum~\cite{Planck:2018vyg}. 
The lensing contribution is determined by the lensing angular power spectrum, which can be rescaled by a parameter $A_{\rm{L}}$,

\begin{equation}
C_{\ell} \to A_{\rm{L}} C_{\ell} .
\end{equation}


Analogously to $A_{\rm{L}}$, we introduce five phenomenological parameters, which correspond to five different physical effects related to CMB angular power spectrum.
Taking both the T-mode and E-mode into account, the photon transfer functions for the scalar modes can be written as follows~\cite{Lesgourgues:2013bra,Howlett:2012mh,Ruiz-Granda:2022bcn},

\begin{eqnarray}
\label{eq:deltat}
  \Delta_{\ell}^{\rm{T}} (k,\eta_0)  &=& A_{\rm{SW}} \int_{0}^{\eta_0} d\eta g(\eta)\[\Delta_0^{\rm{T}}(k,\eta) + \Psi(k,\eta) \] j_{\ell}[k(\eta_0-\eta)] \nonumber\\ 
 &+&A_{\rm{Dop}} \int_{0}^{\eta_0} d\eta g(\eta) \dfrac{\theta_{\rm{b}}}{k}  j'_{\ell}[k(\eta_0-\eta)]  \nonumber \\
  &+&\int_{0}^{\eta_0} d\eta f(\eta) e^{-\tau} \[\Psi'(k,\eta) +\Phi'(k,\eta) \] j_{\ell}[k(\eta_0-\eta)] \nonumber \\
 &+&A_{\rm{Pol}} \int_{0}^{\eta_0} d\eta \dfrac{g(\eta)}{2} P^{(0)} \[ 3j''_{\ell}[k(\eta_0-\eta)\] + j_{\ell}[k(\eta_0-\eta)]] . \\  \nonumber
\end{eqnarray}       
\begin{eqnarray}
\label{eq:deltae}
\Delta_{\ell}^{\rm{E}} (k,\eta_0)  &=&  A_{\rm{Pol}} \int_{0}^{\eta_0} d\eta \sqrt{\dfrac{9(\ell+2)!}{4(\ell-2)!}} g(\eta) P^{(0)} \dfrac{j_{\ell}[k(\eta_0-\eta)]}{[k(\eta_0-\eta)]^2}.                                      
\end{eqnarray}                                                             
$g(\eta)=-\tau' e^{-\tau}$ is the visibility function, where $\tau$ is the optical depth at a given conformal time $\eta$. 
$\Delta_{0}^{\rm{T}}$ represents the intrinsic temperature anisotropy of the monopole. $j_{\ell}[k(\eta_0-\eta)]$ are the spherical Bessel functions, $j'_{\ell}[k(\eta_0-\eta)]$ and $j''_{\ell}[k(\eta_0-\eta)]$ are respectively their first and second derivatives.
$\theta_{\rm{b}}$ denotes the divergence of baryon bulk velocity. $P^{(0)}$ is the scalar polarization source term.  
Additionally,
\begin{equation}
f(\eta) = \begin{cases} 
 A_{\rm{eISW}} ,\ \  \text{for\ } z > 30,  \\  
A_{\rm{lISW}} ,\ \  \text{for\ } z \le30 .
\end{cases} 
\label{eq:fisw}
\end{equation}

From the above formulas, it can be clearly seen the mechanisms of various physical effects and their corresponding rescaling parameters contribute to the amplitude of the T-mode and E-mode. 
For the T-mode, as Eq.~\ref{eq:deltat} shows, the first term incorporates the SW effect, which is rescaled by the phenomenological amplitude $A_{\rm{SW}}$. It is the core contributor to the T-mode and accounts for the temperature anisotropy of CMB photons caused by the gravitational potential at the last scattering surface.
The second term describes the Doppler effect of CMB photons, modulated by a parameter $A_{\rm{Dop}}$. It describes the temperature anisotropy generated by the relative motion when CMB photons scatter with moving baryons, influencing the phase and amplitude of temperature fluctuations.
The third term accounts for the ISW effect, which is the additional temperature anisotropy of CMB photons caused by non-static spacetime evolution and gravitational potential changes during their propagation. We adopt Eq.~\ref{eq:fisw} to describe the eISW and lISW effects, which are parameterized by $A_{\rm{eISW}}$ for early times and $A_{\rm{lISW}}$ for late times respectively. 
Finally, the fourth term represents the contribution of photon polarization and anisotropic Thomson scattering to the angular power spectrum, modulated by the parameter $A_{\rm{Pol}}$.
In Eq.~\ref{eq:deltae}, $A_{\rm{Pol}}$ is used to rescale the E-mode because it is the main manifestation of the polarization effect.
We have to mention that scalar perturbations do not produce the other CMB polarization mode---the B-mode, unless the Lensing effect is considered. That is because the B-mode is generated by the conversion of E-mode through the Lensing effect.

Thus far, we have six phenomenological parameters $A_{\rm{new}}$ (new$=$L, SW, Dop, eISW, lISW, Pol) and all of them have a theoretical expected values of $A_{\rm{new}}=1$ in the $\Lambda$CDM model. 

\begin{figure}[H]
\begin{center}
\includegraphics[width=1\textwidth,height=0.6\textheight]{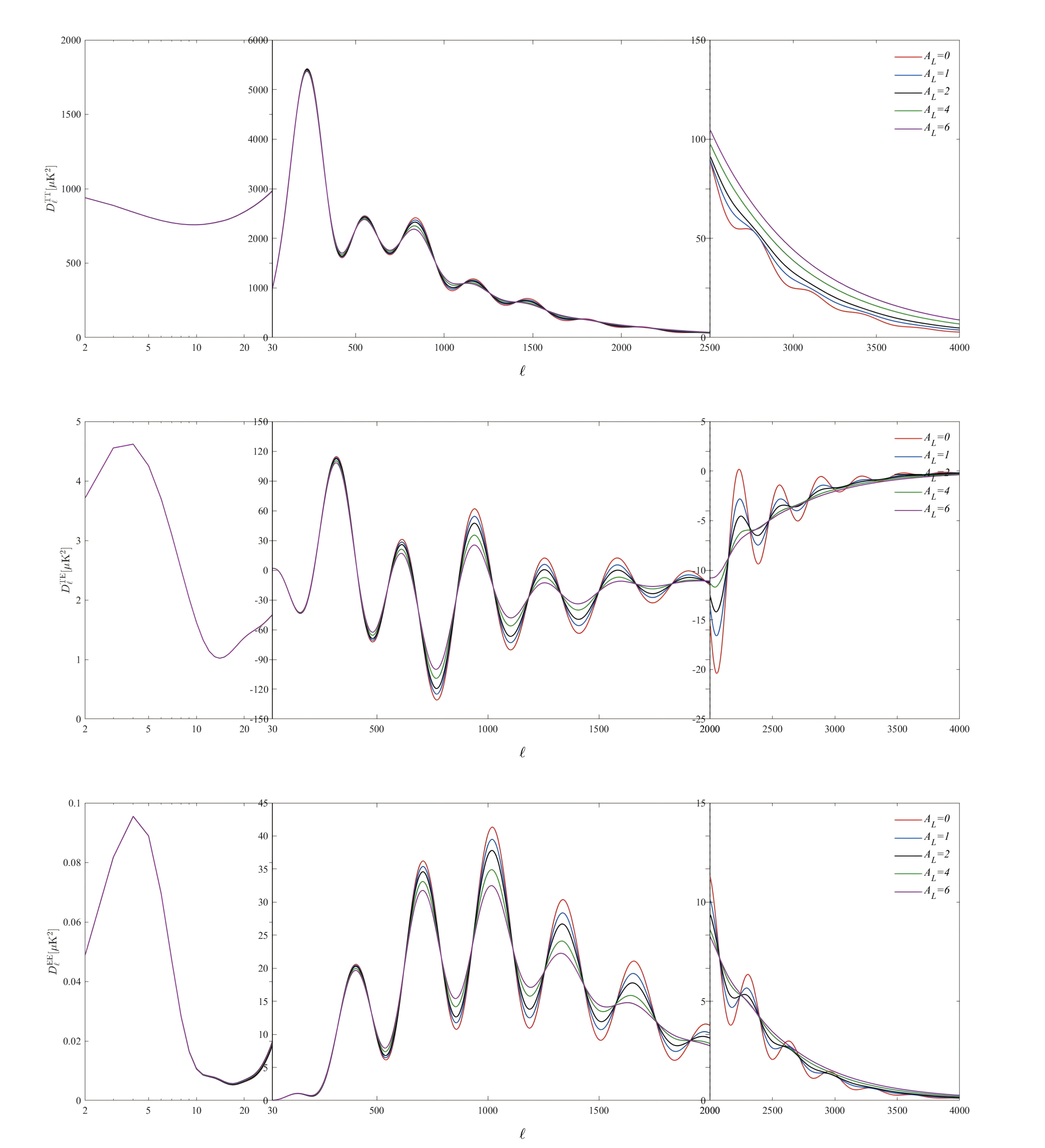}
\end{center}
\caption{The plot shows the impact of changing the physical lensing amplitude $A_{\rm{L}}$ on the theoretical TT, EE, and TE angular power spectra.}
\label{fig:al}
\end{figure}

\begin{figure}[H]
\begin{center}
\includegraphics[width=1\textwidth,height=1\textheight]{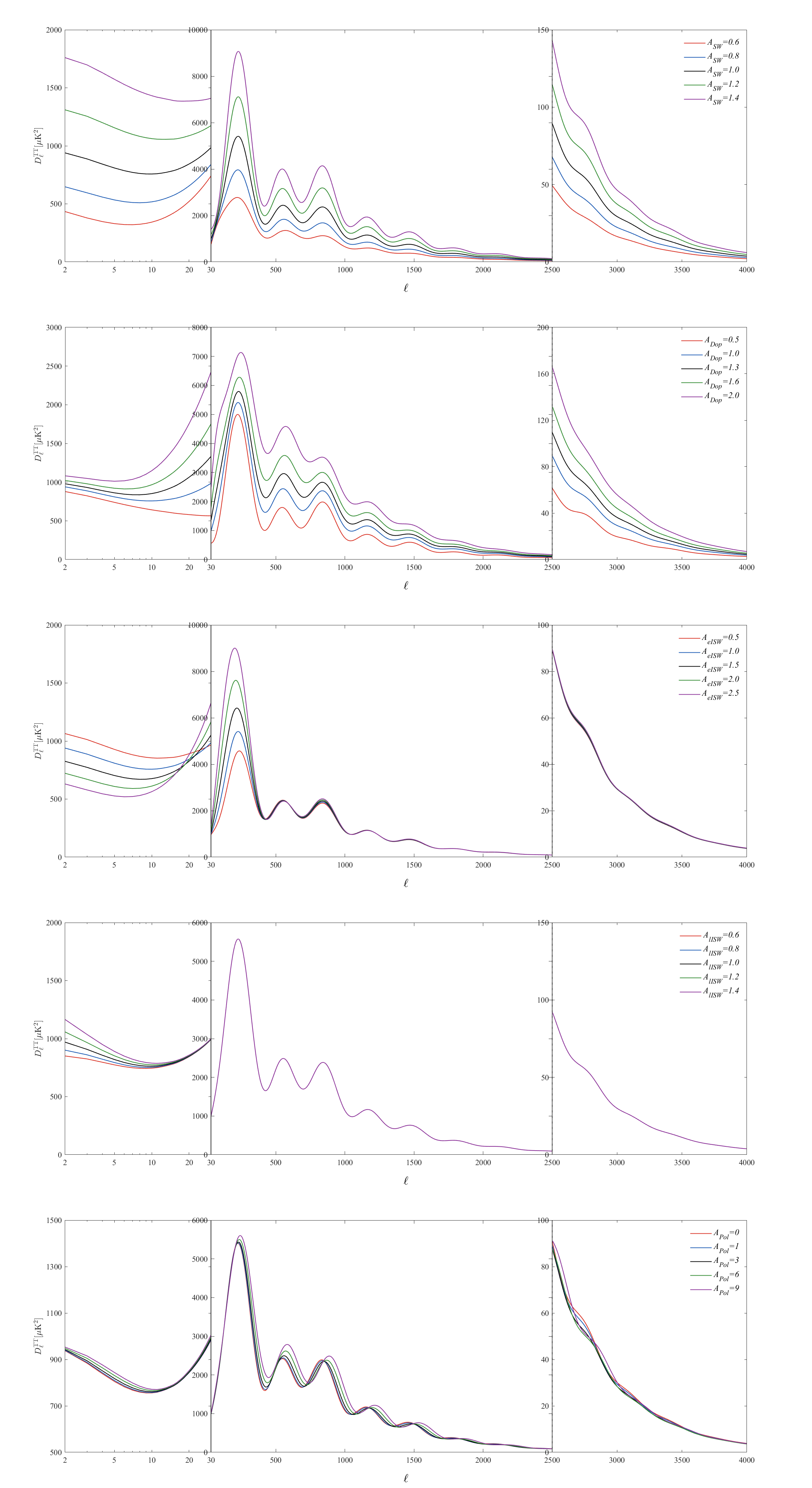}
\end{center}
\caption{The plot shows the impact of changing a single physical amplitude ($A_{\rm{new}}$ excepting $A_{\rm{L}}$) each time on the theoretical TT angular power spectra.}
\label{fig:tt}
\end{figure}

\begin{figure}[H]
\begin{center}
\includegraphics[width=1\textwidth,height=1\textheight]{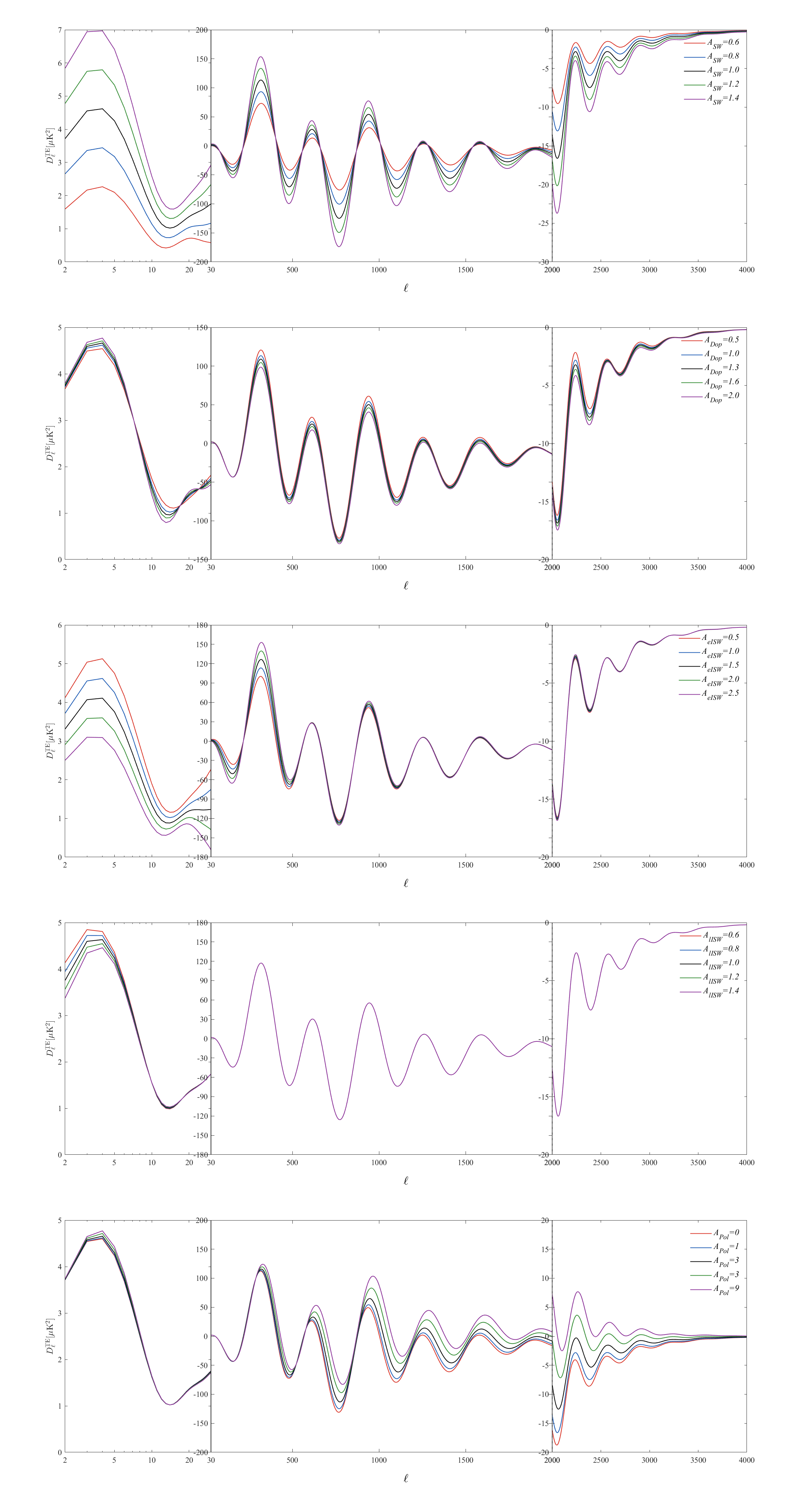}
\end{center}
\caption{The plot shows the impact of changing a single physical amplitude ($A_{\rm{new}}$ excepting $A_{\rm{L}}$) each time on the theoretical TE angular power spectra.}
\label{fig:te}
\end{figure}

To show the influences of above rescaling parameters vividly, we modify the publicly available Boltzmann solver---CAMB package~\cite{Lewis:1999bs,Howlett:2012mh} and plot the TT, TE and EE power spectra with difference values. In Fig.~\ref{fig:al}, we show the influences of varying values of $A_{\rm{L}}$ on the TT, TE and EE angular power spectra. 
Fig.~\ref{fig:tt} and Fig.~\ref{fig:te} illustrate the effects of a single parameter other than $A_{\rm{L}}$ on the TT and TE angular power spectra. 
The selection of different values of $A_{\rm{new}}$ is based on the ability to observe graphical differences and the convenience for comparison with previous studies~\cite{Ruiz-Granda:2022bcn,Kable:2020hcw}. The values of the other necessary parameters are the central values of the $\Lambda$CDM model from Tab.~\ref{tab:plc}.
The vertical axis corresponds to $D_{\ell}=\ell (\ell+1) C_{\ell}/2\pi$ (in unit of $\mu K^2$). And the horizontal axis is $\ell$, ranging from 2 to 4000. The upper limit of $\ell$ is much higher than $\ell < 2000-2500$ in other works because the measurement of ACT DR6~\cite{AtacamaCosmologyTelescope:2025blo} is used in Sec.~\ref{sec:res}. Moreover, we divided the whole range to three bins: 2-30, 30-2000, and 2000-4000, named the low-, mid- or high-$\ell$ bin in this work.
Notice that the figures of EE angular power spectra related to $A_{\rm{new}}$ other than $A_{\rm{L}}$, which resemble those in Fig.~\ref{fig:tt} and Fig.~\ref{fig:te}, are not shown. That is because the figures show negligible variations with the same parameters.

Focusing on Fig.~\ref{fig:al}, the enhancement of the Lensing effect smooths the TT, TE, and EE angular power spectra. As the value of $A_{\rm{L}}$ increases, the peaks decrease and the valleys increase obviously in the mid- and high-$\ell$ bins for all the angular power spectra, but no significant variations are visible in the low-$\ell$ bins.
As shown in Fig.~\ref{fig:tt} and Fig.~\ref{fig:te}, the SW effect has the most significant influence on the angular power spectra compared with other physical effects. 
Increasing $A_{\rm{SW}}$ enhances both the peaks and the valleys in the TT power spectrum in the whole range of $\ell$. For the TE power spectrum, the peaks and valleys increase in the low- and high-$\ell$ bins, but the peaks increase and the valleys decrease in the mid-$\ell$ bin.
The parameter $A_{\rm{Dop}}$ related to the Doppler effect influences the amplitudes of both TT and TE angular power spectra in the whole range of $\ell$, with particularly prominent impact on the second peak of TT angular power spectrum. Moreover, as $A_{\rm{Dop}}$ increases, both the peaks and valleys of TT angular power spectrum increase. Nevertheless, the TE spectrum exhibits a more complex behavior. In the low- and mid-$\ell$ bins, the peaks increase and the valleys decrease as $A_{\rm{Dop}}$ increases, while both the peaks and valleys increase in the high-$\ell$ bin.
Then we find that the eISW effect has influence on the low- and mid-$\ell$ bins, especially $\ell \lesssim 1000$. And it exerts the most pronounced influence on the first peak of the TT and TE angular power spectra. In both cases, an increase in the value of $A_{\rm{eISW}}$ leads to a corresponding rise in the height of the first peak. 
In comparison, $A_{\rm{lISW}}$ has a narrower effective range, limited to the low-$\ell$ bins for both the TT and TE angular power spectra, while negligible variations appear at mid- and high-$\ell$ bins. Specifically, increasing $A_{\rm{lISW}}$ raises the amplitude of the TT angular power spectrum, while reducing that of the TE spectrum.
The Polarization effect causes a phase shift towards the larger $\ell$ in both the TT and TE angular power spectra in the whole range of $\ell$, mainly affecting the second and subsequent peaks. Furthermore, when $A_{\rm{Pol}}$ increases, this effect enhances the TT and TE angular power spectra. 


\section{Datasets, Methodology and Results}
\label{sec:res} 

In the previous section, we define the phenomenological amplitudes of the SW, eISW, lISW, Doppler, Polarization effects, as well as the Lensing effect, which influence the anisotropies in the CMB angular power spectra. In this section, we explore the constraints on these phenomenological amplitudes from different cosmological observations. 

Firstly, we list all the observational data used in our analysis.

\begin{itemize}

\item \textbf{CMB:} 
\begin{itemize}
\item Planck: We use Planck PR3 low-$\ell$ TT Commander likelihood ($2 \leq \ell \leq 29$), Planck PR3 low-$\ell$ SimAll EE likelihood ($2 \leq \ell \leq 29$) \cite{Planck:2018vyg}, the Planck PR4 CamSpec high-$\ell$ temperature and polarization likelihood based on NPIPE maps, covering TTTEEE power spectra from $\ell$ = 30 to $\ell$ = 2500 \cite{Rosenberg:2022sdy, Efstathiou:2019mdh}. 
Furthermore, we incorporate its CMB lensing measurements, which are reconstructed from the CMB lensing potential using Planck PR4 NPIPE maps \cite{Carron:2022eyg}.
\item ACT: We use the ACT DR6 data~\cite{AtacamaCosmologyTelescope:2025blo}, which provides high-resolution ground-based CMB temperature and polarization measurements. Compared to Planck, the dataset cover extensive sky areas and captures information on smaller scales. Concretely, the CMB angular power spectra extend from $\ell=600$ up to $\ell=8500$. Moreover, the lensing measurements from the ACT DR6, which consists of five seasons of CMB temperature and polarization observations, are included in our work~\cite{ACT:2023kun,ACT:2023dou}. 
Notice that we limit the maximum value of $\ell$ to 4000 to mitigate large uncertainties.  
\end{itemize}

\item \textbf{BAO:}
We adopt the BAO measurements from DESI DR2~\cite{DESI:2025zpo,DESI:2025zgx}, obtained from various tracers, including the Bright Galaxy Sample (BGS), Luminous Red Galaxies (LRG), Emission Line Galaxies (ELG), Quasars (QSO), and the Lyman-$\alpha$ (Ly$\alpha$) forests. Compared to previous BAO measurements, the second release of DESI covers a wider range of redshifts $0.1 < z < 4.2$, more samples and higher accuracy. 

\item \textbf{SNIa:} 
As standard candles, the SNIa determine the cosmic distances by using the cosmic distance ladder. 
For this analysis, we utilize the distance moduli measurements of 1701 light curves from 1550 SNIa samples over the redshift range $0.001 < z < 2.26$, which is named PantheonPlus \cite{Scolnic:2021amr, Brout:2022vxf}.

\end{itemize}

In this work, we adopt two data combinations as follows: (i) Planck+DESI+PantheonPlus (PDP); (ii) Planck+ACT+DESI+PantheonPlus (PADP).

We refer to the CAMB package~\cite{Lewis:1999bs,Howlett:2012mh}, and the Markov Chain Monte Carlo (MCMC) sampler---Cobaya~\cite{Torrado:2020dgo} to constrain parameters in the $\Lambda$CDM$+A_{\rm{new}}$ models (new$=$L, SW, Dop, eISW, lISW, Pol), as well as the $\Lambda$CDM model as the base model. 
There are seven free parameters: $\{ \Omega_{\rm{c}} h^2,\ \Omega_{\rm{b}} h^2,\ A_{\rm{s}},\ n_{\rm{s}},\ \tau,\ \theta_{\rm{MC}},\ A_{\rm{new}} \}$. 
Here, $\Omega_{\rm{c}} h^2$ is the physical density of the cold dark matter today, $\Omega_{\rm{b}} h^2$ is the physical density of baryons today, $A_{\rm{s}}$ is the amplitude of the power spectrum of primordial curvature perturbations at the pivot scale $k_{\rm{p}}=0.05 $ Mpc$^{-1}$, $n_{\rm{s}}$ is the scalar spectral index, $\tau$ is the reionization optical depth, $\theta_{\rm{MC}}$ is the ratio between the sound horizon and the angular diameter distance at the decoupling epoch, and $A_{\rm{new}}$ corresponds to the phenomenological amplitude of different physical effects mentioned above.
The convergence of the MCMC chains is evaluated using the Gelman-Rubin parameter $R-1\le 0.01$.

\begin{figure}[H]
\begin{center}
\includegraphics[scale=0.325]{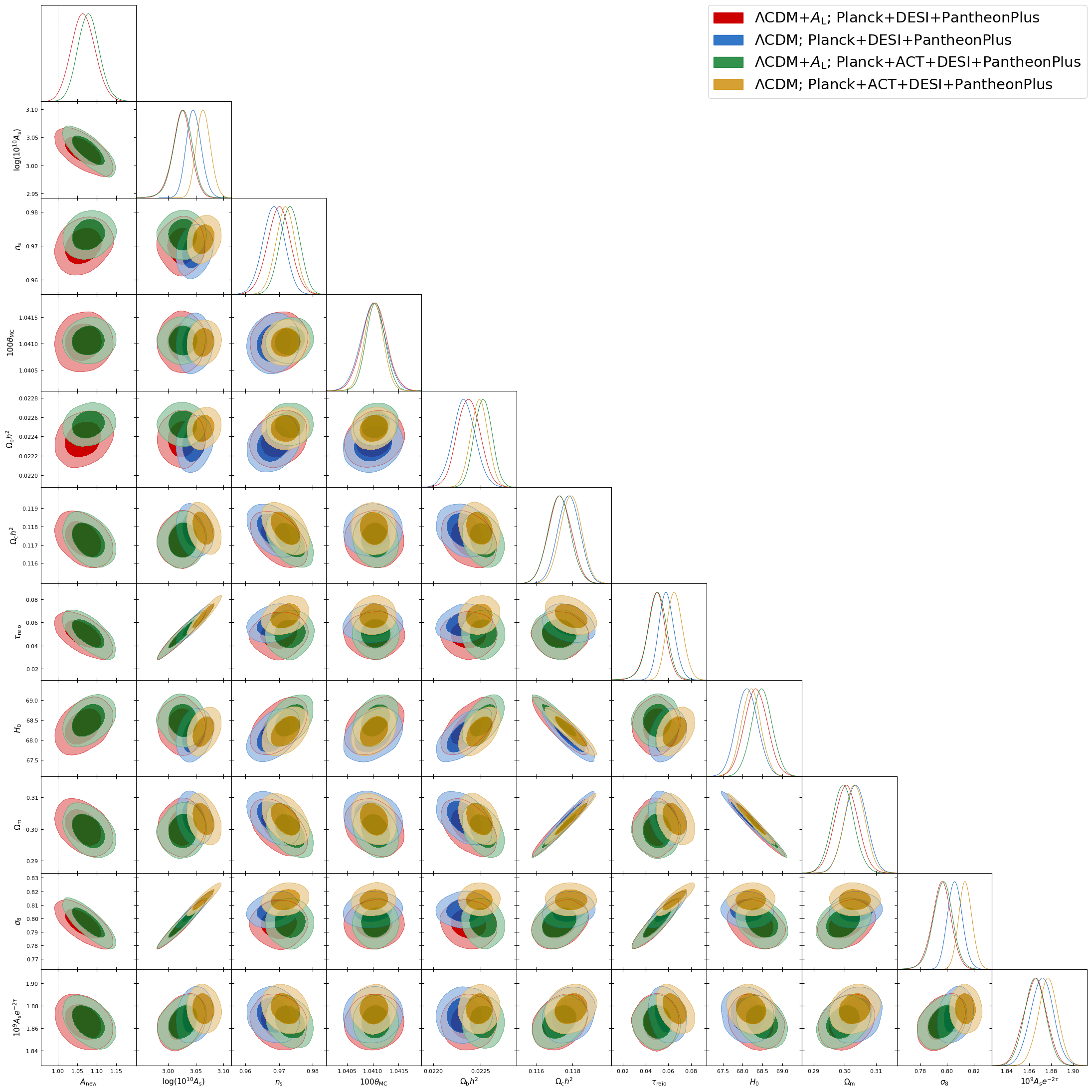}
\end{center}
\caption{The triangular plot of parameters in the $\Lambda$CDM model and $\Lambda$CDM$+A_{\rm{L}}$ model from the PDP (Planck+DESI+PantheonPlus) and PADP (Planck+ACT+DESI+PantheonPlus) datasets.}
\label{fig:tri-1al}
\end{figure}

\begin{figure}[H]
\begin{center}
\includegraphics[scale=0.325]{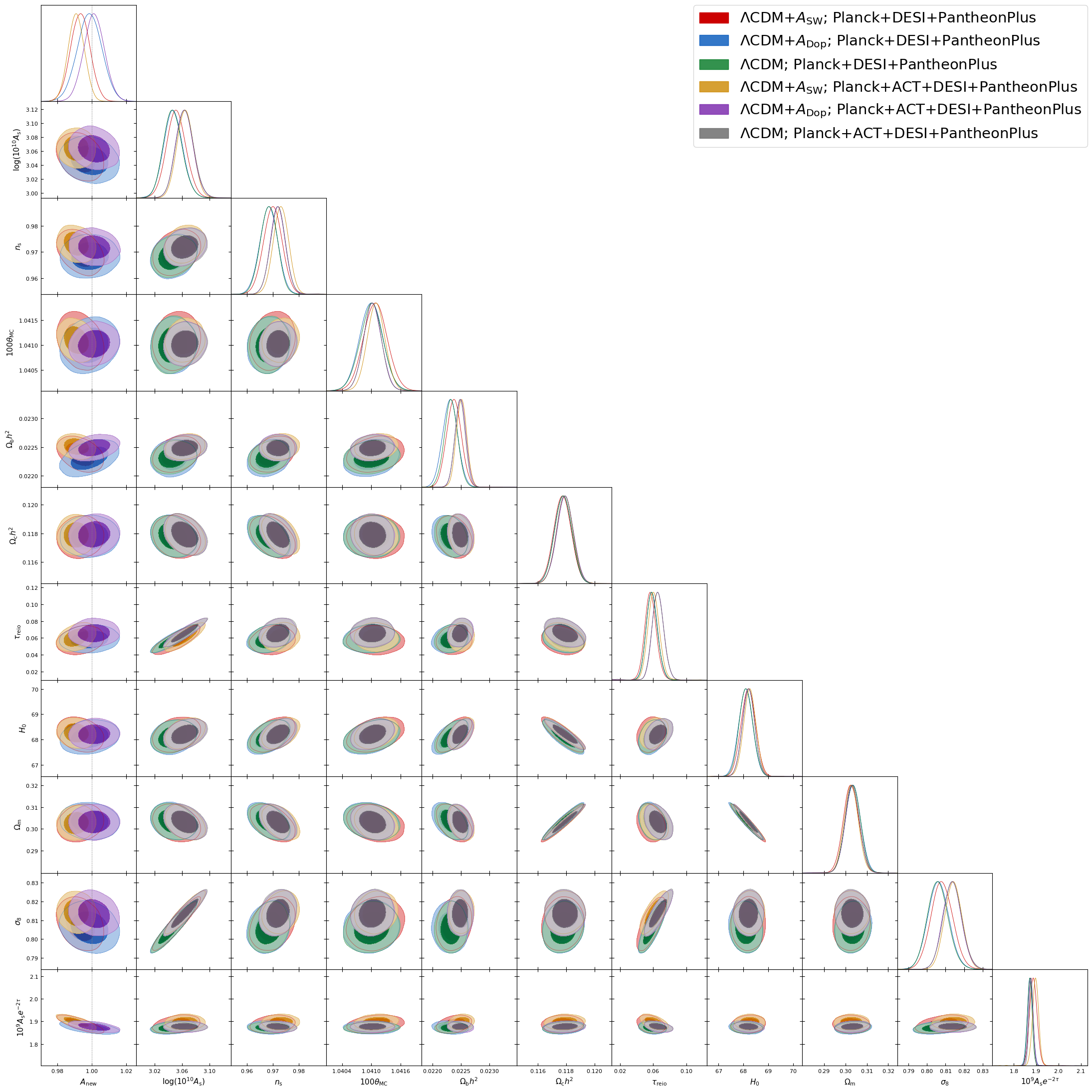}
\end{center}
\caption{The triangular plot of parameters in the $\Lambda$CDM, $\Lambda$CDM$+A_{\rm{SW}}$ and $\Lambda$CDM$+A_{\rm{Dop}}$ models from the PDP (Planck+DESI+PantheonPlus) and PADP (Planck+ACT+DESI+PantheonPlus) datasets.}
\label{fig:tri-2sd}
\end{figure}

\begin{figure}[H]
\begin{center}
\includegraphics[scale=0.325]{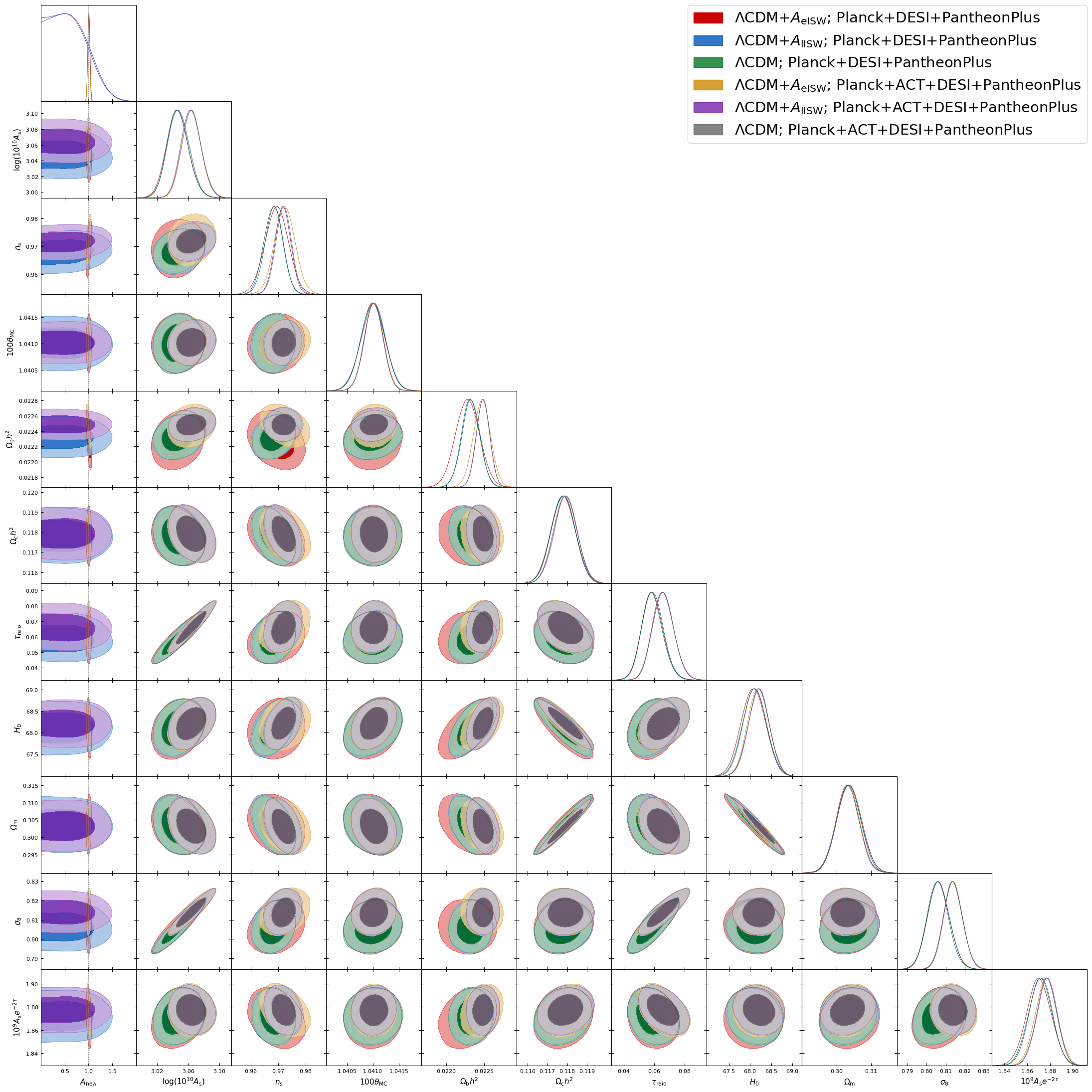}
\end{center}
\caption{The triangular plot of parameters in the $\Lambda$CDM, $\Lambda$CDM$+A_{\rm{eISW}}$ and $\Lambda$CDM$+A_{\rm{lISW}}$ models from the PDP (Planck+DESI+PantheonPlus) and PADP (Planck+ACT+DESI+PantheonPlus) datasets.}
\label{fig:tri-3isw}
\end{figure}

\begin{figure}[H]
\begin{center}
\includegraphics[scale=0.325]{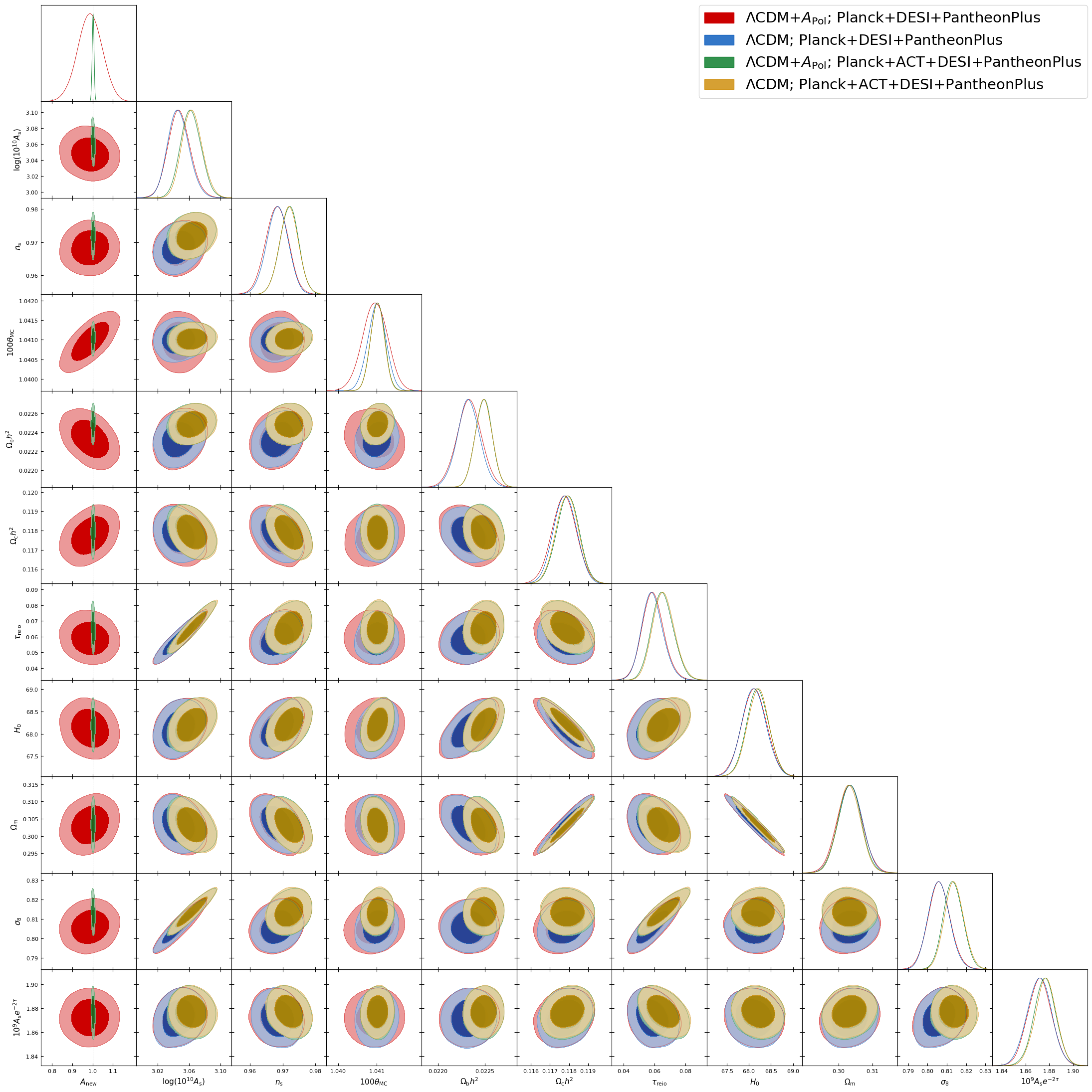}
\end{center}
\caption{The triangular plot of parameters in the $\Lambda$CDM model and $\Lambda$CDM$+A_{\rm{Pol}}$ model from the PDP (Planck+DESI+PantheonPlus) and PADP (Planck+ACT+DESI+PantheonPlus) datasets.}
\label{fig:tri-4pol}
\end{figure}

Tab.~\ref{tab:plc} and Tab.~\ref{tab:plc-act} present the $68\%$ limits for the necessary parameters in the $\Lambda$CDM model and the $\Lambda$CDM$+A_{\rm{new}}$ models from the PDP and PADP data combinations.
Fig.~\ref{fig:tri-1al} - Fig.~\ref{fig:tri-4pol} show the triangular plots of parameters in the $\Lambda$CDM$+A_{\rm{new}}$ models from the previous two datasets, comparing with that of the $\Lambda$CDM model.
To facilitate the analysis, we firstly summarize the free parameters, followed by a discussion on the Hubble tension, $\sigma_8$ tension and the Akaike Information Criterion (AIC) analysis~\cite{Liddle:2007fy,Gong:2007se}.

For the $\Lambda$CDM$+A_{\rm{L}}$ model, we have $A_{\rm{L}}=1.0656^{+0.0304}_{-0.0303}$ at $68\%$ C.L. and $A_{\rm{L}}=1.0656^{+0.0617}_{-0.0587}$ at $95\%$ C.L. from PDP. And $A_{\rm{L}}=1.0795^{+0.0260}_{-0.0289}$ at $68\%$ C.L., $A_{\rm{L}}=1.0795^{+0.0570}_{-0.0532}$ at $95\%$ C.L. from PADP, which deviates over 2$\sigma$ from the standard value $A_{\rm{L}}=1$ in the $\Lambda$CDM model. It implies that the actual gravitational lensing effect may be much stronger than the prediction, which is consistent with Planck Collaboration.  
From Planck PR3, $A_{\rm{L}}>1$ is favored at about 3$\sigma$ when only Planck PR3 TTTEEE$+$lowE are used, and including the Planck lensing reconstruction the result is consistent at 2$\sigma$ with $A_{\rm{L}}=1$~\cite{Planck:2018vyg}.
But there is no evidence for excess lensing effect in the ACT DR6 measurement and $A_{\rm{L}}=1.007\pm 0.057$ at $68\%$ C.L.~\cite{AtacamaCosmologyTelescope:2025blo}. 
Indeed, 
combining it with Planck data only tightens the constraint without pulling the central value closer to 1, owing to the impact of $A_{\rm{L}}$ in the high-$\ell$ bin CMB power spectra.
Four of the free parameters $\{\Omega_{\rm{b}}h^2,\ \Omega_{\rm{c}}h^2,\ n_{\rm{s}},\ 100\theta_{\rm{MC}} \}$ remain essentially unchanged.
In comparison with the $\Lambda$CDM model, with the constraints on $10^9 A_{\rm{s}} e^{-2\tau}$ remaining stable across all models and data combinations, the values of $\log (10^{10}A_{\rm{s}})$ and $\tau$ in the $\Lambda$CDM$+A_{\rm{L}}$ model are reduced, with broader uncertainties. 
From the PDP dataset, $\log (10^{10}A_{\rm{s}}) =3.025^{+0.017}_{-0.016}$ and $\tau=0.0494^{+0.0082}_{-0.0074}$ decreased by approximately 1$\sigma$.
And for the PADP dataset, $\log (10^{10}A_{\rm{s}})=3.026^{+0.018}_{-0.016}$ and $\tau=0.0500^{+0.0086}_{-0.0077}$ show reductions of $1.81\sigma$ and $1.48\sigma$, respectively.
The values are also lower than the results from ACT DR6 which reads $\log (10^{10}A_{\rm{s}})=3.053\pm0.013$ and $\tau=0.0562^{+0.0053}_{-0.0063}$~\cite{AtacamaCosmologyTelescope:2025blo}. 
The results can be intuitively observed from the triangular diagram, as Fig.~\ref{fig:tri-1al} shows.

In the $\Lambda$CDM$+A_{\rm{SW}}$ model, we get $A_{\rm{SW}}=0.9932^{+0.0055}_{-0.0056}$ from PDP and $A_{\rm{SW}}=0.9908^{+0.0048}_{-0.0047}$ from PADP. 
They deviate from the theoretical value of 1 by 1.21$\sigma$ and 1.96$\sigma$ respectively. 
The $\Lambda$CDM$+A_{\rm{Dop}}$ model gives $A_{\rm{Dop}}=0.9987\pm0.0070$ from PDP and $A_{\rm{Dop}}=1.0014\pm0.0061$ from PADP. The standard value $A_{\rm{Dop}}=1$ is still located within the 68$\%$ C.L..
From Planck PR3 TTTEEE$+$lowE$+$lensing data, the obtained values of $A_{\rm{SW}}$ and $A_{\rm{Dop}}$ are consistent with the fiducial value of 1 at 68$\%$ C.L.~\cite{Planck:2018vyg}. As shown in Tab.~\ref{tab:plc} and Tab.~\ref{tab:plc-act}, other parameters of the $\Lambda$CDM$+A_{\rm{SW}}$ model and $\Lambda$CDM$+A_{\rm{Dop}}$ model are consistent with results of the $\Lambda$CDM model. 
These two models present tight and similar constraints, because both $A_{\rm{SW}}$ and $A_{\rm{Dop}}$ are highly sensitive to the CMB angular power spectra over all the angular scales, as shown in Fig.~\ref{fig:tt} and Fig.~\ref{fig:te}.

Then, for the $\Lambda$CDM$+A_{\rm{eISW}}$ model, $A_{\rm{eISW}}=1.0081^{+0.0228}_{-0.0229}$ for PDP, showing no significant deviation from the standard value 1.  
And $A_{\rm{eISW}}=1.0036^{+0.0215}_{-0.0211}$ for PADP, which is slightly but insignificantly larger than 1.
Compared with $A_{\rm{eISW}}=0.991\pm0.026$ from Planck PR3 TTTEEE$+$lowE$+$lensing data, our result decreases the uncertainty by approximately 13$\%$~\cite{Ruiz-Granda:2022bcn}.
For the $\Lambda$CDM$+A_{\rm{lISW}}$ model, $A_{\rm{lISW}}<0.7577$ from PDP, $A_{\rm{lISW}}=0.5989^{+0.1742}_{-0.5963}$ from PADP, which indicates that the dataset cannot effectively constrain this parameter actually. This is reasonable because the impact of $A_{\rm{lISW}}$ on the CMB power spectra is extremely minimal and only exists in the low-$\ell$ bin as shown in Fig.~\ref{fig:tt} and Fig.~\ref{fig:te}. But our constraint on $A_{\rm{lISW}}$ is less stringent than $A_{\rm{lISW}}<0.680$ from Planck PR3 TT$+$lowE and $A_{\rm{lISW}}<0.687$ from Planck PR3 TTTEEE$+$lowE$+$lensing~\cite{Planck:2018vyg}.  

Finally, in the $\Lambda$CDM$+A_{\rm{Pol}}$ model, our results show $A_{\rm{Pol}}=0.9868^{+0.0607}_{-0.0601}$ for PDP and $A_{\rm{Pol}}=1.0026\pm0.0046$ for PADP.
The inclusion of ACT data reduces the uncertainty by more than one order of magnitude. 
This can be attributed to the significant impact of $A_{\rm{Pol}}$ on the high-$\ell$ bin of the TE angular power spectrum shown in Fig.~\ref{fig:te}.
However, Ref.~\cite{Ruiz-Granda:2022bcn} provides $A_{\rm{Pol}}=0.9988\pm0.0026$ from Planck PR3 TTTEEE$+$lowE$+$lensing.
Despite using a more comprehensive dataset, our constraint is not tighter than the Planck-only result in the literature.
Therefore, non-CMB measurements and ACT data do not provide significant improvement in constraining the CMB polarization effect.

We now turn to the Hubble and $\sigma_8$ tensions. 
Across all models, the values of the Hubble constant $H_0 \approx$ 68 km s$^{-1}$ Mpc$^{-1}$ remain essentially stable, indicating that the single-parameter extensions introducing the CMB amplitude parameters $A_{\rm{new}}$ cannot alleviate the Hubble tension. As for the $\sigma_8$ tension, our results are consistent with $\sigma_8=0.8111\pm 0.0060$, $S_8=0.832\pm 0.013$ from Planck~\cite{Planck:2018vyg}. Only the $\Lambda$CDM$+A_\mathrm{L}$ model yields a mild alleviation of the $\sigma_8$ tension. In this scenario, we get $\sigma_8=0.7958^{+0.0073}_{-0.0068}$, $S_8\equiv \sigma_8 \sqrt{\Omega_{\rm{m}}/0.3} =0.7965^{+0.0097}_{-0.0098}$ from PDP, and $\sigma_8=0.7967^{+0.0075}_{-0.0069}$, $S_8=0.7959^{+0.0096}_{-0.0097}$ from PADP, which have 1.53$\sigma$ tension with $S_8=0.766^{+0.020}_{-0.014}$ from KiDS-1000~\cite{Heymans:2020gsg}. 
Collectively, introducing $A_{\rm{new}}$ as an additional parameter to the $\Lambda$CDM model barely relieves the Hubble tension and the $\sigma_8$ tension.

Turning to the AIC analysis, we note that in the $\Lambda$CDM$+A_{\rm{L}}$ model $\Delta$AIC$\equiv$AIC$-$AIC$_{\Lambda \rm{CDM}}=-1.57$ and $-6.11$ for the PDP and PADP datasets, respectively. This indicates that the $\Lambda$CDM$+A_{\rm{L}}$ model is favored by observational data relative to the standard $\Lambda\mathrm{CDM}$ model.
With the inclusion of ACT DR6, the aforementioned tendencies become even more pronounced. This is because the smoothing effect of $A_{\rm{L}}$ on high-$\ell$ data is more significant as shown in Fig.~\ref{fig:al}, and ACT DR6 provides high-resolution measurements at small angular scales. 
For all other considered models, positive $\Delta$AIC values are obtained, implying that these extended scenarios are less supported by the observations. 

 \begin{sidewaystable}
\centering
\renewcommand{\arraystretch}{1.33}
\begin{minipage}{\textwidth}
\centering
\begin{tabular}{|w{c}{2.8cm}|w{c}{3cm}|w{c}{3cm}|w{c}{3cm}|w{c}{3cm}|w{c}{3cm}|w{c}{3cm}|w{c}{3cm}|}
\hline
Parameter                      & $\Lambda$CDM              & $\Lambda$CDM$+A_{\rm{L}}$         &$\Lambda$CDM $+A_{\rm{SW}}$    &$\Lambda$CDM $+A_{\rm{Dop}}$            
   &$\Lambda$CDM $+A_{\rm{eISW}}$   &$\Lambda$CDM $+A_{\rm{lISW}}$       &$\Lambda$CDM $+A_{\rm{Pol}}$\\
\hline
$\Omega_{\mathrm{b}} h^2$      & $0.02232\pm0.00012$       & $0.02237\pm0.00012$               & $0.02238\pm0.00013$            & $0.02231\pm0.00013  $                              
   & $0.02228\pm0.00016$             & $0.02232\pm0.00012$                 & $0.02233\pm0.00013$\\  
$\Omega_{\mathrm{c}} h^2$      & $0.11789^{+0.00061}_{-0.00060}$       &$0.11730^{+0.00064}_{-0.00063}$    & $0.11774\pm0.00061$            & $0.11779\pm 0.00061 $                              
   & $0.11779\pm0.00061$             & $0.11782^{+0.00061}_{-0.00060}$                 & $0.11774\pm0.00064$\\
$\log(10^{10} A_{\mathrm{s}})$ & $3.047^{+0.013}_{-0.014}$ & $3.025^{+0.017}_{-0.016}$         & $3.052^{+0.014}_{-0.015}$      & $3.047\pm0.014 $                                   
   &$3.046\pm0.014$        & $3.047^{+0.013}_{-0.014}$           & $3.048^{+0.013}_{-0.014}$\\
$n_{\mathrm{s}}$               & $0.9684\pm 0.0032$        &$0.9700\pm 0.0034$                 & $0.9700\pm0.0036$   & $0.9684\pm 0.0034$                         
   &$0.9693^{+0.0042}_{-0.0043}$               & $0.9681^{+0.0033}_{-0.0034}$                  & $0.9683\pm 0.0034$  \\
$\tau$                         &$0.0586^{+0.0062}_{-0.0072}$& $0.0494^{+0.0082}_{-0.0074}$     & $0.0568^{+0.0066}_{-0.0072}$   & $0.0585\pm0.0067$                       
   & $0.0588^{+0.0065}_{-0.0072}$    &$0.0589^{+0.0065}_{-0.0072}$         & $0.0590^{+0.0065}_{-0.0072}$\\
$100\theta_{\rm{MC}}$      & $1.04101^{+0.00023}_{-0.00024}$ & $1.04103\pm0.00023$  & $1.04108\pm0.00024 $           & $1.04100^{+0.00023}_{-0.00024}$ & $1.04100\pm0.00023$  & $1.04101^{+0.00022}_{-0.00023}$      & $1.04096\pm0.00032$ \\
$A_{\rm{new}}$                 & -                         & $1.0656^{+0.0304}_{-0.0303}$      &  $0.9932^{+0.0055}_{-0.0056}$             &  $0.9987\pm0.0070$                                 
   & $1.0081^{+0.0228}_{-0.0229}$                      &  $<0.7577$                         &  $0.9868^{+0.0607}_{-0.0601}$\\
\hline
$10^9 A_{\rm{s}} e^{-2\tau}$   & $1.872\pm0.010$           & $1.866\pm0.010$                   & $1.888^{+0.016}_{-0.017}$      & $1.873\pm0.012$                       
   &$1.870\pm0.011$                  & $1.872\pm0.010$                     &$1.872\pm0.010$\\
   
$H_0$[km s$^{-1}$ Mpc$^{-1}$]    & $68.11^{+0.43}_{-0.27}$            & $68.34\pm0.29$                    & $68.20\pm0.29$                 & $68.10\pm0.28$                         
   &$68.08\pm0.29$                   & $68.10\pm0.28$                      & $68.12\pm0.28 $\\
$\Omega_{\rm{m}}$              & $0.3034\pm0.0035$         & $0.3005\pm0.0037$                 & $0.3027\pm0.0036$              & $0.3035\pm0.0035$                                  
   & $0.3036\pm0.0036$               & $0.3036\pm0.0035$                   & $0.3033\pm0.0036$\\
$\sigma_8$                     &$0.8060^{+0.0054}_{-0.0055}$& $0.7958^{+0.0073}_{-0.0068}$     & $0.8081^{+0.0059}_{-0.0058}$              & $0.8062\pm0.0056  $                                
   & $0.8063\pm0.0057$    & $0.8063\pm0.0057$        & $0.8061\pm0.0056$\\
$S_8$                     &$0.8105\pm0.0075$& $0.7965^{+0.0097}_{-0.0098}$     & $0.8116\pm0.0077$              & $0.8109\pm0.0078  $                                
   & $0.8111^{+0.0079}_{-0.0078}$    & $0.8111\pm0.0077$        & $0.8105^{+0.0078}_{-0.0079}$\\
\hline
$\rm{AIC}$                     & $12427.32$                & $12425.75$                        &  $12430.63$                    &  $12431.86$                                        
   &  $12431.80$                     &  $12430.44$                         &  $12435.13$\\
$\rm{AIC}-\rm{AIC_{\Lambda CDM}}$& $0$                     & $-1.57$                           & $3.31$                        & $4.54$                                             
   & $4.48$                           & $3.12$                               & $7.81$\\
\hline
\end{tabular}
\caption{Central values and 68\% limits for the cosmological parameters in the $\Lambda$CDM model and the $\Lambda$CDM$+A_{\rm{new}}$ models from the PDP (Planck+DESI+PantheonPlus) dataset.}
\label{tab:plc}
\end{minipage}
  
\begin{minipage}{\textwidth}
\vspace{20pt}  
\centering
\begin{tabular}{|w{c}{2.8cm}|w{c}{3cm}|w{c}{3cm}|w{c}{3cm}|w{c}{3cm}|w{c}{3cm}|w{c}{3cm}|w{c}{3cm}|}
\hline
Parameter                      & $\Lambda$CDM              & $\Lambda$CDM$+A_{\rm{L}}$         &$\Lambda$CDM$+A_{\rm{SW}}$ &$\Lambda$CDM$+A_{\rm{Dop}}$            &$\Lambda$CDM$+A_{\rm{eISW}}$   &$\Lambda$CDM$+A_{\rm{lISW}}$       &$\Lambda$CDM$+A_{\rm{Pol}}$\\
\hline
$\Omega_{\mathrm{b}} h^2$      & $0.02248\pm0.00009$       & $0.02252\pm0.00009$               & $0.02251\pm0.00009$       & $0.02249\pm0.00009$      & $0.02247\pm0.00011$             & $0.02249\pm0.00009$                 & $0.02249\pm0.00009$\\
$\Omega_{\mathrm{c}} h^2$      & $0.11790\pm0.00058$       &$0.11726^{+0.00060}_{-0.00061}$    & $0.11789^{+0.00056}_{-0.00057}$            & $0.11792^{+0.00059}_{-0.00058} $                              
   & $0.11791^{+0.00058}_{-0.00059}$             & $0.11793\pm0.00058$                                       
  & $0.11794\pm0.00058$\\
$\log(10^{10} A_{\mathrm{s}})$ & $3.064^{+0.012}_{-0.013}$ & $3.026^{+0.018}_{-0.016}$         & $3.065\pm0.012$      & $3.063 ^{+0.012}_{-0.013}$                                   
   &$3.064\pm0.013$        & $3.064\pm0.012$           & $3.062^{+0.012}_{-0.013}$\\
$n_{\mathrm{s}}$               & $0.9720^{+0.0028}_{-0.0029}$        &$0.9732\pm0.0029$                 & $0.9731\pm0.0029$   & $0.9718\pm 0.0029$                         
   &$0.9723^{+0.0038}_{-0.0037}$               & $0.9716\pm0.0028$                  & $0.9720^\pm0.0028$  \\
$\tau$                         &$0.0658^{+0.0065}_{-0.0074}$& $0.0500^{+0.0086}_{-0.0077}$     & $0.0608\pm0.0069$   & $0.0658^{+0.0064}_{-0.0072}$                       
   & $0.0658^{+0.0066}_{-0.0073}$    &$0.0659^{+0.0068}_{-0.0069}$         & $0.0652^{+0.0065}_{-0.0074}$\\
$100\theta_{\mathrm{MC}}$      &$1.04102^{+0.00018}_{-0.00017}$             &$1.04105^{+0.00018}_{-0.00017}$            & $1.04110\pm0.00018 $           & $1.04102\pm0.00018$ 
   &$1.04102\pm0.00018$  &$1.04102^{+0.00017}_{-0.00018}$      & $1.04103^{+0.00018}_{-0.00019}$\\
$A_{\rm{new}}$                 & -                         & $1.0795^{+0.0260}_{-0.0289}$      &  $0.9908^{+0.0048}_{-0.0047}$             &  $1.0014\pm0.0061$                                 
   & $1.0036^{+0.0215}_{-0.0211}$                      &  $0.5989^{+0.1742}_{-0.5963}$                         &  $1.0026\pm0.0046$\\
\hline
$H_0$[km s$^{-1}$ Mpc$^{-1}$]    & $68.21\pm0.25$            & $68.48\pm0.26$                    & $68.26^{+0.25}_{-0.24}$                 & $68.21\pm0.25$                         
   &$68.20\pm0.25$                   & $68.20\pm0.25$                      & $68.20\pm0.25$\\
$\Omega_{\rm{m}}$              & $0.3031\pm0.0033$         & $0.2994\pm0.0034$                 & $0.3027^{+0.0032}_{-0.0033}$              & $0.3032\pm0.0033$                                  
   & $0.3032\pm0.0033$               & $0.3033\pm0.0033$                   & $0.3033\pm0.0033$\\
$\sigma_8$                     &$0.8138\pm0.0049$           & $0.7967^{+0.0075}_{-0.0069}$     & $0.8143\pm0.0046$              & $0.8136^{+0.0049}_{-0.0050} $                                 
   & $0.8139\pm0.0050$    & $0.8140^{+0.0049}_{-0.0048}$        & $0.8133^{+0.0048}_{-0.0052}$\\
$10^9 A_{\rm{s}} e^{-2\tau}$   & $1.877\pm0.009$                                   & $1.866^{+0.010}_{-0.009}$                   & $1.897\pm0.013$      & $1.876\pm0.010$                       
   &$1.876\pm0.009$                  & $1.878\pm0.009$                     &$1.876\pm0.009$\\
$S_8$     &$0.8180^{+0.0066}_{-0.0067}$& $0.7959^{+0.0096}_{-0.0097}$     & $0.8108^{+0.0065}_{-0.0064}$              & $0.8179^{+0.0066}_{-0.0065}$                                
   & $0.8183^{+0.0067}_{-0.0066}$    & $0.8184\pm0.0066$        & $0.8178^{+0.0065}_{-0.0066}$\\
\hline
$\rm{AIC}$                     & $12605.97$                & $12599.86$                        &  $12606.66$                    &  $12610.25$                                        
   &  $12610.36$                     &  $12610.83$                         &  $12613.24$\\
$\rm{AIC}-\rm{AIC_{\Lambda CDM}}$& $0$                     & $-6.11$                           & $0.69$                        & $4.28$                                             
   & $4.39$                           & $4.86$                               & $7.27$\\
\hline
\end{tabular}
\caption{Central values and 68\% limits for the cosmological parameters in the $\Lambda$CDM model and the $\Lambda$CDM$+A_{\rm{new}}$ models from the PADP (Planck+ACT+DESI+PantheonPlus) dataset.}
\label{tab:plc-act}
\end{minipage}
\end{sidewaystable}

\section{Summary}
\label{sec:sum}

This work presents an investigation into six phenomenological amplitude parameters $A_{\rm{new}}$ (new$=$L, SW, Dop, eISW, lISW, Pol) corresponding to the Lensing, Sachs–Wolfe, Doppler, early Integrated Sachs–Wolfe, late Integrated Sachs–Wolfe, and Polarization effects, respectively, that govern the key physical processes underlying CMB anisotropies. 
Employing a modified version of the CAMB Boltzmann code, we analyze the distinct impacts of each parameter on the CMB angular power spectra. 
Then we constrain parameters in the $\Lambda$CDM$+A_{\rm{new}}$ models via the Cobaya package under two data combinations: Planck$+$DESI$+$PantheonPlus (PDP) and Planck$+$ACT$+$DESI$+$PantheonPlus (PADP). 

By assessing the statistical preference via AIC, among all extended models examined, only the $\Lambda$CDM$+A_{\rm{L}}$ model is favored by observational data. 
We obtain $A_{\rm{L}}=1.0656_{-0.0303}^{+0.0304}$ at 68$\%$ C.L. from the data combination of PDP.
Incorporating ACT DR6 data significantly improves the constraint on $A_{\rm{L}}$, with the value of $A_{\rm{L}} = 1.0795_{-0.0289}^{+0.0260}$  deviating from 1 by $3.06\sigma$.
In the $\Lambda$CDM$+A_{\rm{SW}}$ model, the values of $A_{\rm{SW}}$ deviate from 1 by 1.21$\sigma$ and 1.96$\sigma$ for the case of PDP and PADP respectively. 
And $A_{\rm{lISW}}$ remains poorly constrained due to its minimal effect on power spectra at $\ell \geq 30$.
Other phenomenological parameters $A_{\rm{Dop}}$, $A_{\rm{eISW}}$, $A_{\rm{Pol}}$ are found no significant deviation from the fiducial $\Lambda$CDM prediction of 1.
Notably, these single-parameter extended models cannot yield reductions in the Hubble and $\sigma_8$ tensions.

The inclusion of high-$\ell$ CMB data from the ACT DR6 yields improvements to key cosmological constraints to different extents.
Especially, it significantly enhances the statistical preference for the $\Lambda$CDM$+A_{\rm{L}}$ model and reduces the uncertainty of $A_{\rm{Pol}}$ by more than one order of magnitude owing to its sensitivity at high-$\ell$ scales.
These results highlight that high-$\ell$ measurements from ground-based facilities, such as SPT-3G~\cite{SPT-3G:2026krg,SPT-3G:2026mcr,SPT-3G:2025bzu,SPT-3G:2022hvq} and Simons Observatory~\cite{SimonsObservatory:2025wwn,Coppi:2025fmt} data,  provide unique and crucial information for constraining CMB lensing and polarization effects, and should be routinely included in future CMB analyses.

\vspace{5mm}
\noindent {\bf Acknowledgments}
Lu Chen is supported by grants from NSFC (grant No. 12105164). This work has also received funding from project ZR2021QA021 supported by Shandong Provincial Natural Science Foundation and the Youth Innovation Team Plan of Colleges and Universities in Shandong Province (2023KJ350)


\end{document}